\documentclass[12pt]{article}
\usepackage[english]{babel}
\usepackage[cp1251]{inputenc}
\usepackage[a4paper, left=25mm, right=15mm, top=20mm, bottom=20mm]{geometry}
\usepackage{graphicx}
\usepackage{amssymb}
\usepackage{xcolor}
\usepackage{hyperref}
\hypersetup{pdfstartview=FitH, linkcolor=linkcolor,citecolor=citecolor,colorlinks=true} 
\usepackage{amsmath}
\usepackage{setspace}
\usepackage{bm}
\usepackage{cite}
\usepackage{url}
\usepackage[hang,flushmargin]{footmisc} 
\newcommand{\be}{\begin{equation}}
\newcommand{\ee}{\end{equation}}
\renewcommand{\baselinestretch}{1.14}
\date{}
\begin{document}
\definecolor{linkcolor}{HTML}{008000}
\definecolor{citecolor}{HTML}{4682B4}
\large
\title{\bf \LARGE Linearly coupled quantum harmonic oscillators and their quantum entanglement}
\normalsize
\author{D.N. Makarov, K.A. Makarova \\
Northern (Arctic) Federal University, Arkhangelsk, 163002, Russia\\
E-mail: makarovd0608@yandex.ru  }

\maketitle
\begin{abstract}

\begin{spacing}{1}
Quantum harmonic oscillators linearly coupled through coordinates and momenta, represented by the Hamiltonian $ {\hat H}=\sum^2_{i=1}\left( \frac{ {\hat p}^{2}_i}{2 m_i } + \frac{m_i \omega^2_i}{2} x^2_i\right) +{\hat H}_{int} $, where the interaction of two oscillators ${\hat H}_{int} =  i k_1 x_1 { \hat p }_2+ i k_2 x_2 {\hat p}_1 + k_3 x_1 x_2-k_4 {\hat p}_1 {\hat p}_2$, found in many applications of quantum optics, nonlinear physics, molecular chemistry and biophysics. Despite this, there is currently no general solution to the Schr\"{o}dinger equation for such a system. This is especially relevant for quantum entanglement of such a system in quantum optics applications. Here this problem is solved and it is shown that quantum entanglement depends on only one coefficient $R \in (0,1)$, which includes all the parameters of the system under consideration. It has been shown that quantum entanglement can be very large at certain values of this coefficient. The results obtained have a fairly simple analytical form, which facilitates analysis.
\end{spacing}
\end{abstract}

\section{Introduction}
The study of linearly coupled harmonic oscillators through coordinates and momenta is an important direction in modern physics. This interest is primarily due to the fact that models of such systems are found in many applications of quantum optics\cite{Law_1994, Tang_2022, Aspelmeyer_2014}, nonlinear physics \cite{Fano_1957,Fetter_1971,Kim_1989,Han_1990,Iachello_1991,Han_Kim_1999,Bechcicki_2004,Joshi_2010,Paz_2008,Leonardo_2010}, molecular chemistry \cite{Ikeda_1999, Fillaux_2005, Delor_2017} and biophysics \cite{Romero_2014, Fuller_2014, Halpin_2014}. Physical models of coupled harmonic oscillators have been used in many works, for example, the Lie model in quantum field theory \cite{Fetter_1971, Kim_1989, Han_1990} and others. A similar Hamiltonian is also used in biophysics to explain the problem of photosynthesis \cite{Romero_2014, Fuller_2014, Halpin_2014, Leonrdo_2020}.
It has also long been known that in quantum optics a frequency converter, parametric amplifier, Raman and Brillouin scattering, etc. can be described by a coupled harmonic oscillator \cite{Mollow_1967,Louisell_1961,Lu_1974}. Modern research into coupled quantum harmonic oscillators is mainly determined by their quantum entanglement and represents a separate branch of quantum physics. In particular, quantum communication protocols such as quantum cryptography \cite{Ekert_1991}, quantum dense coding \cite{Bennett_1992}, quantum computing algorithms \cite{Shor_1995} and quantum state teleportation \cite{Aspect_1981, Samuel_1998} can be explained with using entangled states. This is due to the fact that such oscillators are a good model of real physical objects. For example, such objects include thermal vibrations of bound atoms, photons in cavities, optical-mechanical cooling, ions in traps, linear beam splitter and much more \cite{Aspelmeyer_2014, Korppi_2018, Makarov_2019_PRA, Makarov_SREP_2021,Makarov_2022_Math_2, Makarov_PRE_2020,Campos_1989}. Coupled harmonic oscillators are one of the main models for studying quantum decoherence (see, e.g. \cite{Galve_2010,Pachon_2015}).

\begin{figure}[h]
\includegraphics[angle=0,width=1\textwidth, keepaspectratio]{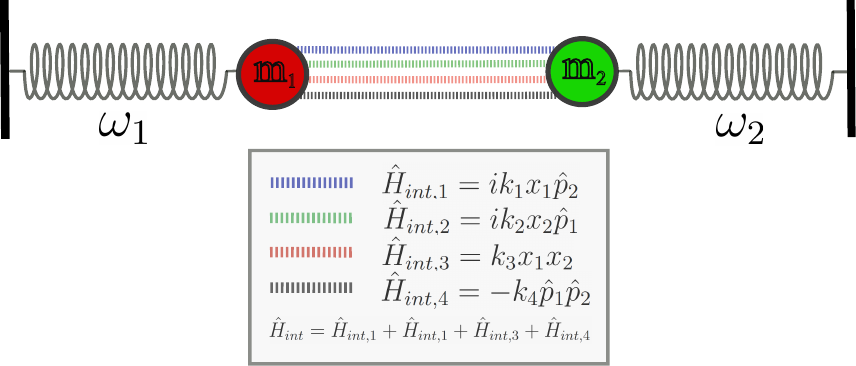}
\caption[fig_1]{The model under study is presented in the form of spring pendulums (oscillators) linearly coupled in 4 different ways through coordinates and momenta.}
\label{fig_1}
\end{figure}

Thus, many systems can be represented as quantum harmonic oscillators linearly coupled in coordinates and momenta with a Hamiltonian in the form
\begin{eqnarray}
{\hat H}=\sum^2_{i=1}\left( \frac{ {\hat p}^{2}_i}{2 m_i } + \frac{m_i \omega^2_i}{2} x^2_i\right) +{\hat H}_{int} ,
\nonumber\\
{\hat H}_{int} = i k_1 x_1 { \hat p }_2+ i k_2 x_2 {\hat p}_1 + k_3 x_1 x_2-k_4 {\hat p}_1 {\hat p}_2, 
\label{1}
\end{eqnarray}
where ${\hat p}_k=-i\hbar \frac{\partial}{\partial x_k}$ ($ k = 1,2 $) is the momentum operator; $m_i, \omega_i$ are the mass and frequency of oscillator $i$, respectively; coefficients $k_1, k_2, k_3, k_4$ determine the coupled between two oscillators with the interaction energy ${\hat H}_{int}$, see Fig.\ref{fig_1}. Often in quantum optics Eq.(\ref{1}) can be seen through the operators of particle creation ${\hat a}_i$ and annihilation ${\hat a}^{\dagger}_i$, in this case the Hamiltonian will be
\begin{eqnarray}
{\hat H}=\sum^2_{i=1}\hbar \omega_i{\hat a}^{\dagger}_i {\hat a}_i + {\hat H}_{int},
\nonumber\\
{\hat H}_{int} = \alpha {\hat a}_1 {\hat a}_2+ \beta {\hat a}_1 {\hat a}^{\dagger}_2+\gamma {\hat a}^{\dagger}_1 {\hat a}_2+\delta {\hat a}^{\dagger}_1 {\hat a}^{\dagger}_2,
\label{2}
\end{eqnarray}
where $\alpha, \beta, \gamma, \delta$ are some constants (below we will show the connection between these constants and the constants Eq.(\ref{1})).
The solution to the Schr\"{o}dinger equation ${\hat H} \Psi = i\hbar \frac{\partial \Psi}{\partial t}$ in general form is currently not known. There are only some special cases of this solution, where the quantum entanglement of such oscillators is also studied. For example, in the works \cite{Han_Kim_1999, Makarov_2018_PRE, Makarov_PRE_2020, Casini_2009} the solution and quantum entanglement of such a system were obtained for $k_1=k_2=k_4=0$, and in the work \cite{Makarov_SREP_2018} a solution was found for $k_2=k_3=k_4=0 $. In the works \cite{Dattoli_1996, Zuniga_2019, Zuniga_2020} the solution for $k_1=k_2=0$ was studied.  Various special cases can be found in the works \cite{Law_1994, Tang_2022, Aspelmeyer_2014} (also see the references of these works).

In order to investigate the solution $|\Psi(x_1,x_2,t)\rangle $ of the Schr\"{o}dinger equation with Hamiltonian Eq.(\ref{1}) and the quantum entanglement of this system, it is necessary not only to find a solution, but also to find the decomposition of this solution into Schmidt modes. It should be added that decomposing the solution into Schmidt modes is a rather difficult task. According to Schmidt's theorem \cite{Ekert_1995,Grobe_1994} the wave function $|\Psi(x_1,x_2,t)\rangle $ of two interacting systems can be expanded as $|\Psi(x_1,x_2,t)\rangle = \sum_{k}\sqrt{\lambda_k (t)}u_{k}(x_1,t)v_{k}(x_2,t) $, where $u_{k}(x_1,t)$ is the wave function of the pure state of the first system, and $v_{k} (x_2,t)$ is the wave function of the pure state of the second system, where $\lambda_k$ is the Schmidt mode. This mode is the eigenvalue of the reduced density matrix, that is, $\rho_1(x_1,x^{'}_1,t)=\sum_{k}\lambda_k(t) u_{k}(x_1,t)u^{* }_{k}(x^{' }_{1},t) $ or $\rho_2(x_2,x^{'}_{2},t)=\sum_{k}\lambda_k(t) v_{k}(x_2,t)v^ {*}_{k}(x^{'}_{2},t) $. To calculate the quantum entanglement of a system, you can use various entanglement measures, for example, the Schmidt parameter \cite{Ekert_1995,Grobe_1994} $K=\left(\sum_{k}\lambda^2_k \right)^{-1} $ or Von Neumann entropy \cite{Bennett_1996,Casini_2009} $S_N=-\sum_{k} \lambda_k \ln \left(\lambda_k \right) $. The main difficulty in calculating quantum entanglement is finding the $\lambda_k$ of the system under consideration. We also add that this work studies quantum entanglement for a dynamic system where the nonstationary Schr\"{o}dinger equation is solved. This means that at the initial moment of time $t = 0$ the system of oscillators was not linearly connected, but at $t>0$ a connection arises in the system as a result of some process (depending on the problem being studied), as a result of which a quantum entanglement of the system.

In this work, a solution to the nonstationary Schr\"{o}dinger equation is found in the analytical form $|\Psi(x_1,x_2,t)\rangle $. It was shown that this solution depends on only two parameters, which include all quantities of the system under consideration. Based on this solution, the Schmidt mode was found and quantum entanglement was studied. It is shown that quantum entanglement depends on just one parameter $R \in (0,1)$, which greatly simplifies the analysis of the problem under consideration. In the resulting expressions, the coefficient $R$ has a simple analytical form and includes all the parameters of the system under consideration. It is shown that for certain values of the coefficient $R$, quantum entanglement can be large.

\section{Solution of the nonstationary Schr\"{o}dinger equation}

To solve Eq.(\ref{1}) it is convenient to go to dimensionless variables; for this we need to replace $\sqrt{\frac{m_i \omega_i}{\hbar}} x_i\to x_i$, we get
\begin{eqnarray}
{\hat H}=\frac{1}{2}\sum^2_{i=1}\hbar \omega_i  \left( -\frac{\partial^2}{\partial x^2_i} +  x^2_i\right) +{\hat H}_{int} ,
\nonumber\\
{\hat H}_{int} = A_{12} x_1 \frac{\partial }{\partial x_2}+ A_{21} x_2 \frac{\partial }{\partial x_1} + B x_1 x_2+C \frac{\partial }{\partial x_1} \frac{\partial }{\partial x_2}, 
\nonumber\\
A_{12}=\hbar k_1 \sqrt{\frac{m_2 \omega_2}{m_1 \omega_1}},~A_{21}=\hbar k_2 \sqrt{\frac{m_1 \omega_1}{m_2 \omega_2}},~B= \frac{\hbar k_3 }{\sqrt{m_1 \omega_1 m_2 \omega_2}},~C=\hbar k_4 \sqrt{m_1 \omega_1 m_2 \omega_2}.
\label{3}
\end{eqnarray}
If we introduce the well-known annihilation operators ${\hat a}_i=\frac{1}{\sqrt{2}}(x_i+\frac{\partial }{\partial x_i})$ and creation ${\hat a}^ {\dagger}_i=\frac{1}{\sqrt{2}}(x_i-\frac{\partial }{\partial x_i})$, then we get Eq.(\ref{2}), where
\begin{eqnarray}
\alpha=\frac{1}{2}\left(A_{12}+A_{21}+B+C\right),~\beta=\frac{1}{2}\left(-A_{12}+A_{21}+B-C\right),
\nonumber\\
\gamma=\frac{1}{2}\left(A_{12}-A_{21}+B-C\right),~\delta=\frac{1}{2}\left(-A_{12}-A_{21}+B+C\right).
\label{4}
\end{eqnarray}
As a result, we need to solve the nonstationary Schr\"{o}dinger equation ${\hat H} \Psi = i\hbar \frac{\partial \Psi}{\partial t}$ with initial conditions $|\Psi(t=0)\rangle = |s_1\rangle |s_2\rangle $ (these states are known solutions of the quantum oscillator),  where $s_1$ and $s_2$ are the quantum numbers of the first and second oscillators before interaction (at ${\hat H}_{int}=0$, i.e. $ {\hat a}^{\dagger}_i {\hat a}_i |\Psi(t=0)\rangle =s_i  \Psi(t=0)\rangle$).

The solution of this kind of differential equations has not previously been found in the literature, with the exception of recently published works \cite{Makarov_RP1_2023, Makarov_RP2_2023, Makarov_Math1_2023}, where it was shown how such an equation can be solved. The standard approach to solving such equations is to diagonalize the Hamiltonian (\ref{3}) by making changes of variables, for example \cite{Makarov_PRE_2020,Makarov_2018_PRE,Han_Kim_1999}. The method of diaganizing the Hamiltonian in the presence of only one of the coefficients $A_{i,j}$ was developed in \cite{Makarov_SREP_2018} by means of a unitary transformation that does not involve a change of variables. To solve Eq. (\ref{3}) we will combine these two methods into one and get a solution. To diagonalize Hamilton's Eq.(\ref{3}), we first make a change of variables in the form $x_1/\sqrt{\hbar \omega_1}=x\cos \theta+y\sin \theta, ~x_2/\sqrt{\hbar \omega_2}= \left(-x\sin \theta+y\cos \theta \right) (1+\delta)$, where $\theta$ and $\delta$ are some unknown coefficients. At the second stage of diaganization, we carry out a unitary transformation over the Hamiltonian, which now depends on the variables $\{x,y\}$, i.e. ${\hat H}={\hat H}(x,y)$ . In other words, we represent $\Psi={\hat S}^{-1}{\Psi}^{'}$, where ${\Psi}^{'}={\hat S}{\Psi}$ . This wave function ${\Psi}^{'}$ will correspond to the Hamiltonian ${\hat H}^{'}={\hat S}{\hat H}(x,y){\hat S}^ {- 1}$ and the conditions ${\hat H}{\Psi}=E{\Psi}$ and ${\hat H}^{'}{\Psi}^{'}=E{\Psi }^ {' }$, where $E$ is the energy eigenvalue. Let us choose a unitary operator ${\hat S}$ in the form ${\hat S}=e^{i\gamma\frac{\partial}{\partial x}\frac{\partial}{\partial y} }e^ {i\alpha x y}$ \cite{Makarov_SREP_2018}, where $\gamma$ and $\alpha $ are some coefficients. Thus, we obtain the Hamiltonian ${\hat H}^{'}$, which has an analytical form, where there are 4 unknown coefficients $\theta, \delta, \gamma, \alpha$. From Eq. (\ref{3}) you can also see that we also have 4 known coefficients $A_{12}, A_{21}, B, C$. As a result, by composing a system of fourth-order equations and setting the coefficients for which there are non-dianalyzable variables equal to zero, we can reduce the Hamiltonian ${\hat H}^{'}$ to a diagonal form. When diaganolizing, we take into account that the oscillators are coupled quite weakly, i.e. coupling coefficients having the dimension of energy (binding energy) will be many times less than the energy of the oscillators $ \{A_{12}, A_{21}, B, C\} \ll \hbar \omega_i$. Here we have described a solution strategy for bringing the Hamiltonian Eq.(\ref{3}) to diagonal form, the main solution is presented in $ \rm \bf Appendix$.
As a result, accurate to the phase, which can be ignored, we obtain the solution
\begin{eqnarray}
|\Psi(x_1,x_2,t)\rangle=\sum^{s_1+s_2}_{n=0} c_{n,s_1+s_2-n} |n \rangle |s_1+s_2-n \rangle, 
\nonumber\\
c_{n,m}=\sum^{s_1+s_2}_{k=0}A^{s_1,s_2}_{k,s_1+s_2-k}A^{*{n,m}}_{k,s_1+s_2-k}e^{-2ik ~ {\arccos\left( \sqrt{1-R} \sin\phi \right)}} ,
\nonumber\\
A^{n,m}_{k,p}=\frac{\mu^{k+n}\sqrt{k!p!}}{(1+\mu^2)^{\frac{s_1+s_2}{2}}\sqrt{n!m!}}P^{(-(1+s_1+s_2), p-n)}_{k}\left(-\frac{2+\mu^2}{\mu^2} \right),
\nonumber\\
\mu =\sqrt{1+\frac{1-R}{R}\cos^2\phi}-\cos\phi\sqrt{\frac{1-R}{R}},
\label{5}
\end{eqnarray}
where $P^{\alpha, \beta}_{\gamma}(x)$ are Jacobi polynomials and the condition $n+m=s_1+s_2$ is satisfied, i.e. the total number of quantum numbers in the system is conserved. States $|n \rangle=C_n e^{-x^2_1/2}H_n(x_1)$ and $|s_1+s_2-n \rangle=C_{s_1+s_2-n} e^{-x^2_2/2}H_{s_1+s_2-n}(x_2)$ ($H_n(x)$ are Hermite polynomials, $C_n$ are known normalization coefficients for a linear oscillator) these are states of oscillators in a state without interaction. So $| c_{n,s_1+s_2-n}|^2$ determines the probability of detecting the system in the states $|n \rangle$ and $|s_1+s_2-n \rangle$. The coefficients $R$ and $\phi$ will be equal
\begin{eqnarray}
R=\frac{\sin^2\left( \Omega t/2 \sqrt{1+\epsilon^2}  \right) }{(1+\epsilon^2)},~\cos\phi=-\epsilon \sqrt{\frac{R}{1-R}},
\nonumber\\
\Omega=\frac{1}{\hbar}\sqrt{|B-C|^2+|A_{12}-A_{21}|^2},~\epsilon=\frac{\omega_2 -\omega_1}{\Omega}. 
\label{6}
\end{eqnarray}
From the Eq.(\ref{5}) one can see that $\lambda_n=|c_{n,s_1+s_2-n}|^2$ is the Schmidt mode in the Schmidt expansion. You can see that the entire dependence in the wave function is reduced to two variables: $R$ and $\phi$. The properties of the resulting expressions are such that when calculating $\lambda_n=|c_{n,s_1+s_2-n}|^2$ the dependence on $\phi$ disappears and the entire dependence is reduced to only one parameter $R$. Since quantum entanglement is calculated through the Schmidt mode $\lambda_n$, then quantum entanglement depends only on $R$. This remarkable result makes it possible to analyze probability and quantum entanglement quite simply. In addition, you can see that $R \in (0,1)$ is a certain parameter characterizing the degree of interaction of two oscillators. The results obtained coincide with previously known special cases, for example, in the works \cite{Han_Kim_1999, Makarov_2018_PRE, Makarov_PRE_2020, Casini_2009} the result was obtained for $A_{12}=A_{21}=C=0$, and in the work \cite{ Makarov_SREP_2018} at $A_{21}=B=C=0$. From Eq.(\ref{6}) we can see that when $A_{12}=A_{21}$ and $B=C$ the frequency $\Omega =0$, which will lead to $R=0$, i.e. no interaction. This result is quite interesting in that there is in fact an interaction, but it is compensated by the actions of the various members of ${\hat H}_{int}$. This in turn suggests that by varying the parameters of the system $A_{12}, A_{21}, B, C$ one can control the interaction of the oscillators and even consider the interaction to be zero, without breaking the coupling between the oscillators, i.e. the coupling parameters can be non-zero. Also from Eq. (\ref{6}) one can see that the dynamics of the system will be noticeable for $\omega_2 -\omega_1 \lesssim \Omega$, i.e. the frequencies $\omega_1$ and $\omega_2$ should be quite close, since we considered $ \{A_{12}, A_{21}, B, C\} \ll \hbar \omega_i$. 

It should be added that the parameter $\Omega$ in the case of writing through the interaction parameters $\alpha, \beta, \gamma, \delta$, i.e. if we represent the Hamiltonian in terms of Eq.(\ref{2}), it will be $\Omega=\frac{\sqrt{2}}{\hbar}\sqrt{|\beta|^2+|\gamma|^2}$. The result is quite strange at first glance, since the dependence on two parameters $\alpha, \delta$ has disappeared. In fact, this should be the case, since we believe that $ \{A_{12}, A_{21}, B, C\} \ll \hbar \omega_i$, and in this case, as it turns out strictly mathematically, are preserved quantum numbers before and during interaction (or they often talk about conservation of the number of particles). Conservation of quantum numbers means that the interaction occurs in such a way that the creation and annihilation operators must be in the combinations $ {\hat a}_1 {\hat a}^{\dagger}_2$ or $ {\hat a}^{\dagger }_1 {\hat a}_2$. This means that one operator of the first oscillator annihilation a state, and another operator of the second oscillator creates a state instead of the annihilation one, and vice versa. Only in this case will quantum states be preserved. Another interesting conclusion can also be drawn. Since $ \{\alpha, \beta, \gamma, \delta \} \ll \hbar \omega_i$, then the operators for the coefficients $\alpha$ and $\delta$, respectively ${\hat a}_1 {\hat a}_2$ and ${\hat a}^{\dagger}_1 {\hat a}^{\dagger}_2$ are operators of higher order of smallness than the operators $ {\hat a}_1 {\hat a}^ {\dagger}_2$ and $ {\hat a}^{\dagger }_1 {\hat a}_2$, even in order of comparable parameters $ \alpha, \beta, \gamma, \delta$.

\section{Quantum entanglement of oscillators}
Let's consider quantum entanglement based on two measures this is the Schmidt parameter  $K=\left(\sum^{s_1+s_2}_{n=0}\lambda^2_n \right)^{-1} $ \cite{Ekert_1995,Grobe_1994} and Von Neumann entropy $S_N=-\sum^{s_1+s_2}_{n=0} \lambda_n \ln \lambda_n   $ \cite{Bennett_1996 ,Casini_2009}. To calculate these measures of quantum entanglement, you need to directly use Eq.(\ref{5}) with $\lambda_n=|c_{n,s_1+s_2-n}|^2$. Let us present here only some particular values of quantum entanglement in the initial states of the oscillators $S_N(|s_1\rangle ,|s_2\rangle )$ and $K(|s_1\rangle ,|s_2\rangle )$. As an example, let's imagine:
\begin{itemize}
\item for $ s_1 = 1 $ and $ s_2 = 1 $ 
\begin{eqnarray}
S_N (|1\rangle,|1\rangle)=-(1-2R)^2\ln (1-2R)^2-4R(1-R)\ln \left( 2R(1-R)\right),
\nonumber\\
K(|1\rangle,|1\rangle)=\frac{1}{1-8R(1-R)(1-3R(1-R))}.
\label{7_1}
\end{eqnarray}
The maximum quantum entanglement will be $[S_N(|1\rangle,|1\rangle)]_{max}=\ln 3$ and $[K(|1\rangle,|1\rangle)]_{max}=3$ for $R=1/2(1\pm 1/\sqrt{3})$.
\item for $ s_1 = 0 $ and $ s_2 = 2 $ (similarly for $ s_1 = 2 $ and $ s_2 = 0 $) 
\begin{eqnarray}
S_N (|0\rangle,|2\rangle)=2 (-1 + R) (R \ln 2 + \ln(1 - R)) - 2 R \ln R,
\nonumber\\
K(|0\rangle,|2\rangle)=\frac{1}{1-2R(1-R)(2-3R(1-R))}.
\label{7_2}
\end{eqnarray}
The maximum quantum entanglement will be $[S_N(|0\rangle,|2\rangle)]_{max}= \frac{3}{2} \ln 2$ and $[K(|0\rangle,|2\rangle)]_{max}=8/3$ for $R=1/2$.
\item for $ s_1 = 2 $ and $ s_2 = 2 $ 
\begin{eqnarray}
S_N (|2\rangle,|2\rangle)=12 (1 - 2 R)^2 (-1 + R) R \ln\left(-6 (1 - 2 R)^2 (-1 + R) R\right) - 
\nonumber\\
 12 (-1 + R)^2 R^2 \ln\left(6 (-1 + R)^2 R^2\right) - (1 + 6 (-1 + R) R)^2 \ln\left((1 + 6 (-1 + R) R)^2\right),
\nonumber\\
K(|2\rangle,|2\rangle)=\frac{1}{1-24R(1-R)\Bigl(1-3R(1-R)\bigl(4-5R(1-R)(4-7R(1-R))\bigr)\Bigr)}.
\label{7_3}
\end{eqnarray}
The maximum quantum entanglement will be $[S_N(|2\rangle,|2\rangle)]_{max}=1.5381$ for $R=1/2(1\pm 0.3675)$ and $[K(|2\rangle,|2\rangle)]_{max}=4.4312$ for $R=1/2(1\pm 0.3898)$.
\end{itemize}
Quantum entanglement of some special cases can be found in general form, for example, for the state $|s_1\rangle,|0\rangle$ (or similarly for $|0\rangle,|s_2\rangle$). Using the results of the work \cite{Makarov_PRE_2020} for the state $|s\rangle,|0\rangle$ we get
\begin{eqnarray}
K(|s\rangle,|0\rangle)= \frac{1}{(1-R)^{2s} {_2F}_1\left(-s,-s;1;\left( \frac{R}{1-R} \right)^2\right)} ,
\label{8}
\end{eqnarray}
where ${_2F}_1(a,b;c;x)$ is Gaussian hypergeometric function. Also, analyzing the Eq. (\ref {8}), you can get that the maximum of this function at $ R = 1/2 $. With this value of $ R = 1/2 $, one can obtain a simpler expression for quantum entanglement 
\begin{eqnarray}
[K(|s\rangle,|0\rangle)]_{max}= 2^{2s}\frac{(s!)^2}{(2 s)!}.
\label{9}
\end{eqnarray}
Can also be found from Eq. (\ref{9}) parameter $K$ for large values of the quantum number $s$, where we obtain $K_{max}(s\gg1)\to \sqrt {\pi s}$. It can be seen that in this case quantum entanglement is unlimited from above.

There is also another important case, when $s_1=s_2=s$ (for even values $s$) and R=1/2. In this case, Holland-Burnett (HB)\cite{Holland_1993} states are realized. It is well known that this wave function is of great interest in various fields of physics, for example, in quantum metrology \cite{Polino_2020,Pezze_2018}. In this case we get
\begin{eqnarray}
K(|s\rangle,|s\rangle)=\frac{\pi (s!)^2}{\Gamma(s+1/2)^2 {_{4}}F_{3}(1/2,1/2,-s,-s;1,1/2-s,1/2-s;1)} ,
\label{10}
\end{eqnarray}
where $\Gamma(x)$ is the gamma function, ${_{4}}F_{3}(x_1,x_2,x_3,x_4;y_1,y_2,y_3;1)$ is the generalized hypergeometric function. It should be added that the Eq.(\ref{10}) has a fairly simple approximation for sufficiently large $s$; we obtain it in the form $ K = s^{0.897} $.

\begin{figure}[h]
\includegraphics[angle=0,width=1\textwidth, keepaspectratio]{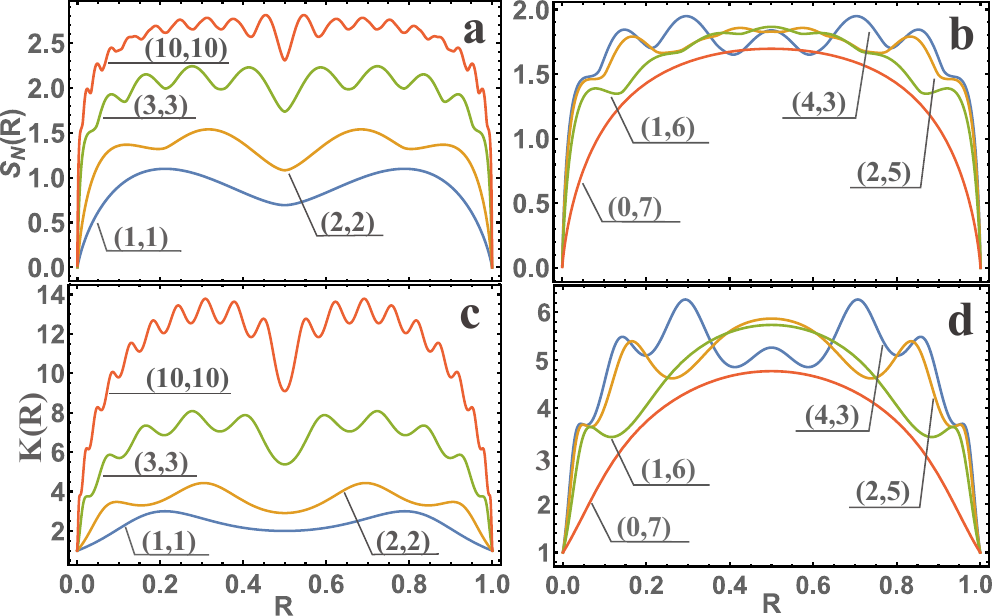}
\caption[fig_2]{The dependence of the Von-Neumann entropy $S_N$ in figures (a) and (b), as well as the Schmidt parameter in figures (c) and (d) as a function of $R$ is presented. In the figures, the dependencies are presented for different initial values of quantum numbers ($s_1, s_2$). For example, when $s_1=1$ and $s_2=9$, the notation (1,9) is entered.}
\label{fig_2}
\end{figure}

Below we present the results of calculations of quantum entanglement for the Von-Neumann entropy $S_{N}$ and the Schmidt parameter $K$ depending on the parameter $R\in (0,1)$, see Fig.\ref{fig_2}.
From these figures and their analysis of the obtained Eqs.(\ref{7_1})-(\ref{10}) conclusions can be drawn. The larger the initial quantum numbers $s_1$ and $s_2$, the greater the quantum entanglement. For the same sum of quantum numbers $s_1+s_2$, the quantum entanglement does not differ much from each other, although for $s_1=s_2$ the quantum entanglement is the highest for certain values of the parameter $R$. Quantum entanglement is a symmetric quantity with respect to $R=1/2$, i.e. quantum entanglement is the same for $R=1/2\pm \Delta R$. It can be seen that quantum entanglement has an oscillatory form with respect to $R$ and the larger $s_1+s_2$, the greater the number of oscillations.

\section{Conclusion}
Thus, in this work, a solution was found for quantum harmonic oscillators Eq.(\ref{5}), linearly related through coordinates and momenta. This solution has an analytical form and allows one to calculate the probabilities of finding oscillators in different states $P_n=|c_{n,s_1+s_2-n}|^2$. Also, the probability depends on just one parameter of the system $R$, see Eq.(\ref{6}), which greatly simplifies the analysis of the resulting expressions. The Schmidt expansion and the Schmidt parameter $\lambda_n=|c_{n,s_1+s_2-n}|^2$ were found, which allows one to calculate the quantum entanglement of a system depending on just one parameter $R$. It was shown that the number of quantum states is conserved before interaction, during interaction and after, i.e. $s_1+s_2=n+m$. Some special cases of quantum entanglement were considered, where the result can be represented in a simple analytical form, and conclusions were drawn. It should be added that quantum entanglement was obtained without taking into account the environment, i.e. external environment. In this case, the environment can significantly affect quantum entanglement\cite{Leonard0_2019} and this is a separate topic for study that is not within the scope of this work.

\small
\begin{spacing}{0.4}

\end{spacing}

\normalsize
\section*{Acknowledgements}

The study was supported by the Russian Science Foundation, project No. 20-72-10151.

\section*{Appendix}
\def\theequation{Ap.\arabic{equation}}
\setcounter{equation}{0}

Let us pose the problem of diaganalyzing the Hamiltonian Eq.(\ref{3}). We make the change of variables $x_1/\sqrt{\hbar \omega_1}=x\cos \theta+y\sin \theta,~x_2/\sqrt{\hbar \omega_2}=\left(-x\sin \theta+y\cos \theta \right) (1+\delta)$,  where $\theta$ and $\delta$ are some unknown coefficients. Next, it is convenient to go to the system of units, where $\hbar=1$. As a result, we get the Hamiltonian
\begin{eqnarray}
{\hat H} = \frac{1}{2}\left(\omega^{2}_{1,x} x^2-a\frac{\partial^2}{\partial x^2}\right)+\frac{1}{2}\left(\omega^{2}_{2,y} y^2-b\frac{\partial^2}{\partial y^2}\right)+
\sum^2_{i,j=1}A^{'}_{i,j} y_i \frac{\partial}{\partial y_j} + B^{'} x y  + C^{'} \frac{\partial}{\partial x}\frac{\partial}{\partial y},
\nonumber\\
\omega^{ 2}_{1,x}=\omega^{2}_1 \cos^2 \theta+\omega^{2}_2 (1+\delta)^2 \sin^2 \theta - B \sqrt{\omega_1\omega_2} \sin 2\theta (1+\delta),
\nonumber\\
\omega^{ 2}_{2,y}=\omega^{2}_2 \cos^2 \theta (1+\delta)^2 +\omega^{2}_1 \sin^2 \theta + B \sqrt{\omega_1\omega_2} \sin 2\theta (1+\delta),
\nonumber\\
A^{'}_{1,1}=-\frac{1}{2}\sin 2\theta \left(\frac{A_{1,2}}{1+\delta}\sqrt{\frac{\omega_1}{\omega_2}}+\sqrt{\frac{\omega_2}{\omega_1}}A_{2,1}(1+\delta)\right),
\nonumber\\
A^{'}_{2,2}=\frac{1}{2}\sin 2\theta \left(\frac{A_{1,2}}{1+\delta}\sqrt{\frac{\omega_1}{\omega_2}}+\sqrt{\frac{\omega_2}{\omega_1}}A_{2,1}(1+\delta)\right),
\nonumber\\
A^{'}_{1,2}=\frac{A_{1,2}}{1+\delta}\sqrt{\frac{\omega_1}{\omega_2}}\cos^2\theta-\sqrt{\frac{\omega_2}{\omega_1}}A_{2,1}(1+\delta)\sin^2\theta,
\nonumber\\
A^{'}_{2,1}=-\frac{A_{1,2}}{1+\delta}\sqrt{\frac{\omega_1}{\omega_2}}\sin^2\theta+\sqrt{\frac{\omega_2}{\omega_1}}A_{2,1}(1+\delta)\cos^2\theta,
\nonumber\\
B^{'}=\frac{1 }{2}\sin 2\theta \left(\omega^{2}_1-\omega^{2}_2(1+\delta)^2\right)+ B \sqrt{\omega_1\omega_2}\left(\cos^2\theta(1+\delta)-\sin^2\theta (1+\delta) \right),
\nonumber\\
C^{'}=-\frac{\sin 2\theta}{2}+\frac{\sin 2\theta}{2(1+\delta)^2}+\frac{C}{\sqrt{{\omega_1}{\omega_2}}}\frac{\cos 2\theta}{1+\delta},
\nonumber\\
a=\cos^2\theta+\frac{\sin^2\theta}{(1+\delta)^2}+\frac{\sin 2\theta}{1+\delta}\frac{C}{\sqrt{{\omega_1}{\omega_2}}},~b=\sin^2\theta+\frac{\cos^2\theta}{(1+\delta)^2}-\frac{\sin 2\theta}{1+\delta}\frac{C}{\sqrt{{\omega_1}{\omega_2}}},
\label{1B}
\end{eqnarray}
where $y_1=x,y_2=y$.

We need to find a solution to the Schr\"{o}dinger equation ${\hat H}{\Psi}=i\frac{\partial \Psi}{\partial t}$, where the Hamiltonian ${\hat H}$ is determined from Eq.(\ref{1B}). Next, we perform a unitary transformation over the desired wave function $\Psi={\hat S}^{-1}{\Psi}^{'}$, where ${\Psi}^{'}={\hat S}{\Psi}$. This wave function ${\Psi}^{'}$ will correspond to the Hamiltonian ${\hat H}^{'}={\hat S}{\hat H}(x,y){\hat S}^ {- 1}$, and the conditions ${\hat H}{\Psi}=E{\Psi}$ and ${\hat H}^{'}{\Psi}^{'}=E{\Psi }^ {'}$, where $E$ is the energy eigenvalue. We choose the unitary operator ${\hat S}$ in the form ${\hat S}=e^{i\gamma\frac{\partial}{\partial x}\frac{\partial}{\partial y} }e^ {i\alpha x y}$, where $\gamma$ and $\alpha $ are some coefficients. To carry out such calculations, we use the well-known expansion
\begin{equation}
e^{\hat X}\hat Y e^{-\hat X} = \hat Y + \left[\hat X, \hat Y\right] + \frac{1}{2!}\left[\hat X, \left[\hat X, \hat Y\right]\right] + \frac{1}{3!}\left[\hat X, \left[\hat X, \left[\hat X, \hat Y\right]\right]\right] + \ldots \ .
\nonumber
\end{equation}
Having carried out all the calculations, we can see that the Hamiltonian ${\hat H}^{'}$ has a finite form (the action of the operators gives a zero value at the 3rd stage).  As a result, the Hamiltonian ${\hat H}^{'}$ can be reduced to a diagonal form (under the condition $ \{A_{12}, A_{21}, B, C\} \ll  \omega_i$) if the unknown coefficients are (for convenience, we redesignate $\theta\to \theta_1$)
\begin{eqnarray}
\alpha=\epsilon_1-\frac{\epsilon_1}{|\epsilon_1|}\sqrt{1+\epsilon^2_1},~\gamma=\frac{\epsilon_1}{2|\epsilon_1|}\frac{1}{\sqrt{1+\epsilon^2_1}},~\tan 2\theta_1= \frac{|B-C|}{ \Delta \omega},~\delta=\frac{C}{ \omega_1}\cot 2\theta_1,
\nonumber\\
\epsilon_1=\frac{ \Delta \omega}{\cos 2\theta_1 |A_{12}-A_{21}|},~ \Delta \omega = \omega_2-\omega_1 .
\label{2B}
\end{eqnarray}
It should be added that the obtained coefficients in Eq.(\ref{2B}) make sense only when $\Delta \omega \lesssim \{A_{12}, A_{21}, B, C\}$, otherwise the system will be in its original state, i.e. does not evolve. Passing for convenience to the dimensionless variables $\{x,y\}$ (this is also taken into account in the coefficients $\alpha$ and $\beta$ in Eq.(\ref{2B})), we obtain the Hamiltonian in the diagonal form
\begin{eqnarray}
{\hat H}^{'}=\frac{\Omega_1}{2}\left(x^2-\frac{\partial^2}{\partial x^2} \right)+\frac{\Omega_2}{2}\left(y^2-\frac{\partial^2}{\partial y^2} \right)+(A^{'}_{1,1}-i\alpha C )x\frac{\partial}{\partial x}+(A^{'}_{2,2}-i\alpha C )y\frac{\partial}{\partial y},
\nonumber\\
\Omega_1=\sqrt{\Omega^2_0+\sigma},~\Omega_2=\sqrt{\Omega^2_0-\sigma},~\sigma= i \omega_1 \frac{\epsilon_1}{|\epsilon_1|}\sqrt{1+\epsilon^2_1} (A^{'}_{1,2}-A^{'}_{2,1}),
\nonumber\\
\Omega^2_0=\omega^{2}_0 a b+i \omega_0 \epsilon_1 (A^{'}_{1,2}+A^{'}_{2,1}),~\omega_0 =i\sqrt{\frac{A^{'}_{2,1}\omega^{2}_{1,x}-A^{'}_{1,2}\omega^{2}_{2,y}}{a A^{'}_{1,2}-bA^{'}_{2,1}}}.
\label{3B}
\end{eqnarray}
Considering that $ \{A_{12}, A_{21}, B, C\} \ll \omega_i$ it is not difficult to obtain
\begin{eqnarray}
\Psi^{'}_k(x)=C_k e^{-x^2/2}H_k(x),~\Psi^{'}_p (y)=C_p e^{-y^2/2}H_p(y),
\nonumber\\
E_k=\Omega_1 \left(k+\frac{1}{2}\right)+c_1,~ E_p=\Omega_2 \left( p+\frac{1}{2}\right)+c_2,
\label{4B}
\end{eqnarray}
where $H_k(x)$ are Hermite polynomials, $c_1$ and $c_2$ non-essential constants, $E_k$ and $E_p$ energy. Find the total energy $E_{k,p}=E_k+E_p$.  We take into account that $\sigma/\Omega^2_0\ll 1$. Expanding in a series in terms of this small parameter and discarding constant values (which do not affect the quantities under study), we get $E_{k,p}=\Omega_0 (k+p)+\frac{\sigma}{2 \omega_1}(k-p)$ (here we took into account that $\Omega_0=\omega_1$ up to an expansion term which can be neglected). Further we will use this energy $E_{k,p}$, although, as will be shown below, the law of conservation of quantum numbers will be satisfied $k+p=const$, which means that the first term in the energy is a constant value and its can be ignored. Let's consider the parameter $\frac{\sigma}{\omega_1}$. It is easy to show that it will be equal to
\begin{eqnarray}
\frac{\sigma}{\omega_1}=\Omega \sqrt{1+\epsilon^2},~\Omega=\sqrt{|B-C|^2+|A_{12}-A_{21}|^2},~\epsilon=\frac{\Delta \omega}{\Omega}.
\label{5B}
\end{eqnarray}

As a result, the general solution of our problem, without choosing the initial conditions, will look like
\begin{eqnarray}
\Psi^{'}(x,y,t)=\sum_{k,p} A_{k,p}\Psi^{'}_k(x)\Psi^{'}_p(y) e^{-iE_{k,p} t},
\label{6B}
\end{eqnarray}
where $A_{k,p}$ are expansion coefficients. To find $\Psi(x,y,t)=e^ {-i\alpha x y} e^{-i\gamma\frac{\partial}{\partial x}\frac{\partial}{\partial y } } \Psi^{'}(x,y,t)$. As a result, we get
\begin{eqnarray}
\Psi(x,y,t)=\sum_{k,p} A_{k,p}e^{-iE_{k,p} t} \Psi_{k,p}(x,y), ~ \Psi_{k,p}(x,y)=e^ {-i\alpha x y} e^{-i\gamma\frac{\partial}{\partial x}\frac{\partial}{\partial y } } \Psi^{'}_k(x)\Psi^{'}_p(y),
\nonumber\\
\Psi(x_1,x_2,t)=\Psi(x,y,t),~x=x_1\cos \theta - x_2 \sin \theta,~ y= x_1\sin \theta + x_2 \cos \theta.
\label{7B}
\end{eqnarray}
Also, the wave function $\Psi(x_1,x_2,t)$ can be expanded in terms of eigenfunctions of the noninteracting system $|\Psi(x_1,x_2,t)\rangle=\sum_{n,m}c_{n,m}|n\rangle | m\rangle e^{-i\varepsilon_{n,m} t}$, where $p_{n,m}=|c_{n,m}|^2$ is the probability of detecting the first and second oscillator in states with $n$ and $m$ quantum numbers, respectively. Using equation (\ref{7B}) it can be shown thatt
\begin{eqnarray}
c_{n,m}=\sum_{k,p} A^{s_1,s_2}_{k,p}A^{*{n,m}}_{k,p}e^{-i E_{k,p}t},~  A^{s_1,s_2}_{k,p}=\langle \Psi_{k,p}(x_1,x_2)|s_1,s_2\rangle,
\label{8B}
\end{eqnarray}
where $ |s_1,s_2\rangle = |\Psi(x_1,x_2,t=0)\rangle $, and $s_1,s_2$ are the quantum numbers of the first and second oscillator before interaction, respectively, i.e. quantum numbers of oscillators in the initial state.

Next, we calculate $A^{s_1,s_2}_{k,p}$ in equation (\ref{8B}) analytically. The expression $\Psi_{k,p}(x,y)$ can be represented not as the action of an operator on $\Psi^{'}_{k}(x)\Psi^{'}_{p}( y)$ , but in the form of an integral expression. To do this, we need to represent $\Psi^{'}_{k}(x)$ through the Fourier integral, i.e. $\Psi^{'}_{k}(x)=\frac{(-i )^n}{\sqrt{2 \pi}}C_n \int^{\infty}_{-\infty} e^ {-{\frac{p^2}{2}}} H_n(p) e^ {i p x } dp$. As a result we get
\begin{eqnarray}
\Psi_{k,p}(x,y)=e^ {-i\alpha x y} e^{-i\gamma\frac{\partial}{\partial x}\frac{\partial}{\partial y } } \Psi^{'}_k(x)\Psi^{'}_p(y) = 
\nonumber\\
\frac{(-i)^nC_n C_m}{\sqrt{2\pi}\sqrt{1+\alpha \gamma}} \int^{\infty}_{-\infty} e^ {-{\frac{p^2}{2}}} H_n(p) e^{i  x \Bigr(\frac{p}{\sqrt{1+\alpha \gamma}}-\alpha y \Bigr)} e^{-\frac{1}{1+\alpha \gamma} \Bigl(y+\frac{\gamma p}{\sqrt{1+\alpha \gamma}} \Bigr)^2} H_m\Bigl( \frac{y+\frac{\gamma p}{\sqrt{1+\alpha \gamma}}  }{\sqrt{1+\alpha \gamma}} \Bigr) dp  .
\label{1C}
\end{eqnarray}
It can be seen that the function $\Psi_{k,p}(x,y)$ is representable only in integral form, and the Fourier transform $\Psi_{k,p}(p,y)=\frac{1}{\sqrt{2 \pi}} \int^{\infty}_{-\infty} \Psi_{k,p}(x,y) e^{ip x} dx$ of it is an analytic function, we obtain
\begin{eqnarray}
\Psi_{k,p}(p,y)=C_k C_p i^{n} e^{-\frac{\xi}{2} (p-\alpha y)^2 }H_k \Bigl( \sqrt{\xi}(p-\alpha y)\Bigr)e^{-\frac{\xi}{2} (y+\alpha p)^2 }H_p \Bigl( \sqrt{\xi}(y+\alpha p)\Bigr),~\xi=\frac{1}{1+\alpha^2}.
\label{2C}
\end{eqnarray}
The coefficient $A^{s_1,s_2}_{k,p}$ in Eq.(\ref{8B}) can be calculated in another way using Eq.(\ref{2C}). For this, we note that in Eq.(\ref{7B})
\begin{eqnarray}
\Psi(p,y,t)=\frac{1}{\sqrt{2\pi}}\int^{\infty}_{-\infty} \Psi(x,y,t)e^{ip x} dx=\sum_{k,p} A_{k,p}e^{-iE_{k,p} t} \Psi_{k,p}(p,y). 
\label{3C}
\end{eqnarray}
From Eq.(\ref{3C}) one can see (similarly to Eq.(\ref{8B})) that
\begin{eqnarray}
c_{n,m}=\sum_{k,p} A^{s_1,s_2}_{k,p}A^{*{n,m}}_{k,p}e^{-i E_{k,p}t},~  A^{s_1,s_2}_{k,p}=\langle \Psi_{k,p}(p,y)|\Psi(p,y,t=0)\rangle.
\label{4C}
\end{eqnarray}
Find $\Psi(p,y,t=0)$ from initial conditions
\begin{eqnarray}
\Psi(p,y,t=0)=\frac{1}{\sqrt{2\pi}}\int^{\infty}_{-\infty} \Psi(x,y,0)e^{ip x} dx=\sum_{k_1,p_1} (-i)^{k_1}B^{s_1,s_2}_{k_1,p_1}(\theta_1) |k_1,p_1 \rangle,  
\label{5C}
\end{eqnarray}
where $|k_1,p_1 \rangle=|k_1\rangle p_1 \rangle$ are states of non-interacting oscillators, and $|k_1\rangle$ depends on the variable $p$, $B^{s_1,s_2}_{k_1,p_1}(\theta_1) =\langle \Psi_{k,p}(x,y) |s_1,s_2 \rangle$ (of course considering that $x=x_1\cos \theta_1 - x_2 \sin \theta_1,~ y= x_1\sin \theta_1 + x_2 \cos \theta_1$). 

Next, you can see that the function $\Psi_{k,p}(p,y)=C_k C_p i^{k}e^{-\frac{p{'}^{2}}{2}}H_k(p{'})e^{-\frac{y{'}^{2}}{2}}H_p(y{'})$ has exactly the same structure as $\Psi_{k,p}(x,y)$ if you notice that $p^{'}=p\cos \theta_2 - y \sin \theta_2,~ y^{'}= y \cos \theta_2 + p \sin \theta_2$ ($\tan \theta_2 = \alpha$). As a result, then we get by substituting Eq.(\ref{5C}) into Eq.(\ref{4C}) (for clarity, let's add $A^{s_1,s_2}_{k,p}=A^{s_1,s_2}_{k,p}(\Theta)$)
\begin{eqnarray}
A^{s_1,s_2}_{k,p}(\Theta)=\sum_{k_1,p_1} (-i)^{k_1}B^{s_1,s_2}_{k_1,p_1}(\theta_1)B^{* k,p}_{k_1,p_1}(\theta_2) .
\label{4CC}
\end{eqnarray}
The properties of the coefficient $B^{n,m}_{k,p}$ have been well studied before, see eg. \cite{Makarov_PRE_2020, Makarov_2018_PRE, Makarov_SREP_2018} and it is equal to
\begin{eqnarray}
B^{n,m}_{k,p}(\theta)=\frac{\mu^{k+n}\sqrt{k!p!}}{(1+\mu^2)^{\frac{s_1+s_2}{2}}\sqrt{n!m!}}P^{(-(1+s_1+s_2), p-n)}_{k}\left(-\frac{2+\mu^2}{\mu^2} \right),~ \mu =\tan \theta,
\label{5CC}
\end{eqnarray}
where $P^{\alpha, \beta}_{\gamma}(x)$ are Jacobi polynomials and the condition $n+m=k+p$ is satisfied. Based on the properties of this coefficient, we can immediately say that the total number of quantum numbers will be conserved in the interaction $s_1+s_2=n+m$. This is an important conclusion of this theory. Further, it can be shown that there is a certain relation between the two angles $\theta_1$ and $\theta_2$, viz
\begin{eqnarray}
\epsilon=\frac{\epsilon_1 \epsilon_2}{\sqrt{1+\epsilon^2_1+\epsilon^2_2}},~\tan 2\theta_1=\frac{1}{\epsilon_1},~\tan 2\theta_2=\frac{1}{\epsilon_2}.
\label{6C}
\end{eqnarray}
Using the properties of the Jacobi polynomials and Eq.(\ref{6C}), one can show that the angle $\Theta $ in Eq.(\ref{4C}) satisfies the condition $\tan 2 \Theta=1/\epsilon$. Moreover, it turns out that the coefficients $B^{n,m}_{k,p}(\theta)=A^{n,m}_{k,p}(\theta)$. As a result, we get that
\begin{eqnarray}
A^{n,m}_{k,p}(\Theta)=\frac{\mu^{k+n}\sqrt{k!p!}}{(1+\mu^2)^{\frac{s_1+s_2}{2}}\sqrt{n!m!}}P^{(-(1+s_1+s_2), p-n)}_{k}\left(-\frac{2+\mu^2}{\mu^2} \right),~ \mu =\tan \Theta.
\label{5CCC}
\end{eqnarray}
It was shown in \cite{Makarov_PRE_2020} that Eq.(\ref{4C}) can be represented in a more convenient form by expressing it in terms of the reflection coefficient and the phase shift.

As a result, we can find the probability of detecting the system in the final states $n$ and $m$ in the first and second oscillators, respectively, when the system transitions from the initial state $s_1,s_2$ in the form $P_n=\left|c_{n,s_1+s_2-n}\right|^2 $, where
\begin{eqnarray}
c_{n,m}=\sum^{s_1+s_2}_{k=0}A^{s_1,s_2}_{k,s_1+s_2-k}A^{*{n,m}}_{k,s_1+s_2-k}e^{-2ik ~ {\arccos\left( \sqrt{1-R} \sin\phi \right)}} ,
\nonumber\\
A^{n,m}_{k,p}=\frac{\mu^{k+n}\sqrt{k!p!}}{(1+\mu^2)^{\frac{s_1+s_2}{2}}\sqrt{n!m!}}P^{(-(1+s_1+s_2), p-n)}_{k}\left(-\frac{2+\mu^2}{\mu^2} \right),
\nonumber\\
\mu =\sqrt{1+\frac{1-R}{R}\cos^2\phi}-\cos\phi\sqrt{\frac{1-R}{R}},
\label{51}
\end{eqnarray}
where $P^{\alpha, \beta}_{\gamma}(x)$ are Jacobi polynomials and the condition $n+m=s_1+s_2$ is satisfied, i.e. the total number of quantum numbers in the system is conserved. Coefficients $R$ and $\phi$ have sense of reflection coefficients, and $\phi$ - phases which will be equal
\begin{eqnarray}
R=\frac{\sin^2\left( \Omega t/2 \sqrt{1+\epsilon^2}  \right) }{(1+\epsilon^2)},~\cos\phi=-\epsilon \sqrt{\frac{R}{1-R}},
\nonumber\\
\Omega=\sqrt{|B-C|^2+|A_{12}-A_{21}|^2},~\epsilon=\frac{\omega_2 -\omega_1}{\Omega}.
\label{61}
\end{eqnarray}

\end{document}